\documentclass[aps,prl,twocolumn,numerical,superscriptaddress,nofootinbib,showpacs,showkeys,longbibliography]{revtex4-1}

\usepackage{graphicx,color}
\usepackage{amsmath,amssymb,amsfonts,bm}
\usepackage[colorlinks=true,linkcolor=blue,plainpages=false]{hyperref}

\newcommand{\be}{\begin{equation}}
\newcommand{\ee}{\end{equation}}
\newcommand{\ba}{\begin{eqnarray}}
\newcommand{\ea}{\end{eqnarray}}
\newcommand{\ban}{\begin{eqnarray*}}
\newcommand{\ean}{\end{eqnarray*}}

\def\v2{\mbox{$v_2$}}

\begin{document}

\title{Signatures of Chiral Magnetic Effect in the Collisions of Isobars}
\medskip

\author{Shuzhe Shi}
\affiliation{Department of Physics, McGill University, 3600 University Street, Montreal, QC, H3A 2T8, Canada.} 

\author{Hui Zhang}
\affiliation{Institute of Particle Physics (IOPP) and Key Laboratory of Quark and Lepton Physics (MOE),  Central China Normal University, Wuhan 430079, China.
} 
\affiliation{Guangdong Provincial Key Laboratory of Nuclear Science, Institute of Quantum Matter, South China Normal University, Guangzhou 510006, China.}

\author{Defu Hou}
\email{houdf@mail.ccnu.edu.cn}
\affiliation{Institute of Particle Physics (IOPP) and Key Laboratory of Quark and Lepton Physics (MOE),  Central China Normal University, Wuhan 430079, China.
} 

\author{Jinfeng Liao}
\email{liaoji@indiana.edu}
\affiliation{ Physics Department and Center for Exploration of Energy and Matter,
Indiana University, 2401 N Milo B. Sampson Lane, Bloomington, IN 47408, USA.} 

 \date{\today}


\begin{abstract}
Quantum anomaly is a fundamental feature of chiral fermions. In chiral materials the microscopic anomaly leads to nontrivial macroscopic transport processes such as the Chiral Magnetic Effect (CME), which has been in the spotlight lately across disciplines of physics. The quark-gluon plasma (QGP) created in relativistic nuclear collisions provides the unique example of a chiral material consisting of  intrinsically relativistic chiral fermions. Potential discovery of CME in QGP is of utmost significance, with  extensive experimental searches  carried out over the past decade. A decisive new collider experiment, dedicated to detecting CME in the collisions of isobars, was performed in 2018 with analysis now underway. In this paper, we develop the state-of-the-art theoretical tool for describing CME phenomenon in these collisions and propose an appropriate isobar subtraction strategy for best background removal. Based on that, we make quantitative predictions for signatures of CME in the collisions of isobars. A new and robust observable that is independent of axial charge uncertainty --- the ratio between isobar-subtracted $\gamma-$ and $\delta-$ correlators, is found to be $- ( 0.41 \pm  0.27 )$  for event-plane measurement and   $ - (  0.90 \pm 0.45 )$ for reaction-plane measurement. 
\end{abstract}

\maketitle


{\em Introduction.---} The investigation of novel quantum transport in chiral materials is a rapidly growing   area of research that has attracted significant interests and activities recently from  a broad range of  physics disciplines such as high energy physics, condensed matter physics, astrophysics, cold atomic gases, etc. Chiral materials are many-body quantum systems that consist of massless fermions (i.e. chiral fermions) that are either fundamental particles or emergent quasi-particles behaving as chiral fermions. A notable example of the former, is the so-called quark-gluon plasma (QGP) which is a new phase of hadronic matter existing at primordially high temperatures available in the early Universe and which is now recreated in laboratories by high energy nuclear collisions~\cite{Kharzeev:2015znc,Miransky:2015ava,Fukushima:2018grm,Liao:2014ava,Hattori:2016emy,Bzdak:2019pkr}. The novel examples of the latter include the latest discovered topological phases of condensed matter systems known as Dirac and Weyl semimetals~\cite{Armitage:2017cjs,Yan:2017jgt,Hasan:2017hwf,Burkov:2017rgl}.

The most salient  feature of chiral fermions is the chiral anomaly under the presence of gauge interactions. Chiral materials manifest such {\em microscopic quantum peculiarity}  through unique {\em macroscopic anomalous transport processes}, which are forbidden in normal environment yet become possible (and necessary) in such chiral materials. A famous example  is the so-called Chiral Magnetic Effect (CME), predicting the generation of an electric current in  chiral materials as response to an applied magnetic field. The CME is a remarkable example as a new kind of quantum electricity that one may call ``magnetricity''. The observation of CME in various physical systems is of fundamental importance.  In semimetal systems the CME-induced transport has been measured via  observables like negative magnetoresistance~\cite{Son:2012bg,Li:2014bha,PhysRevX.5.031023,2016NatCo...711615A}. In the subatomic chiral material, i.e. the quark-gluon plasma created in relativistic nuclear collisions, enthusiastic efforts have been made  to look for its evidence  at  Relativistic Heavy Ion Collider (RHIC) and  Large Hadron Collider (LHC)~\cite{Kharzeev:2007jp,Fukushima:2008xe,Abelev:2009ac,Abelev:2009ad,Adamczyk:2014mzf,Abelev:2012pa,Acharya:2017fau,Khachatryan:2016got,Sirunyan:2017quh}. Despite many measurements  accumulated so far from RHIC and LHC with encouraging hints, the interpretation of these data remains inconclusive due to background contamination. 
The main challenge is that the flow-driven background contribution is proportional to elliptic flow and mimicking the desired CME signal, as clearly revealed by event-shape analysis in e.g. \cite{Acharya:2017fau,Sirunyan:2017quh}.
See discussions in e.g.~\cite{Kharzeev:2015znc,Bzdak:2019pkr,Bzdak:2012ia,Zhao:2019hta}.

An unambiguous observation of CME   in the subatomic system would  be its  first confirmation  in a chiral material of intrinsic relativistic fermions. Its detection would also    provide   tantalizing experimental verification for  the high-temperature restoration of a spontaneously broken global symmetry (the chiral symmetry) which is a fundamental  prediction of the Quantum Chromodynamics (QCD).  It would additionally open a unique window for characterizing the intriguing topological fluctuations of gluon fields --- the non-Abelian gauge fields of QCD. Given such importance,  a decisive  isobaric collision experiment has been carried out in 2018 at RHIC, with the dedicated goal of discovering the CME~\cite{Voloshin:2010ut,Skokov:2016yrj,Kharzeev:2019zgg}.  The basic idea, as initially suggested by Voloshin in \cite{Voloshin:2010ut}, is to contrast the CME-sensitive observables in two different colliding systems, the RuRu and the ZrZr, where the Ru and Zr are a pair of isobar nuclei with the same nucleon numbers ($A=96$) but different nuclear charges ($Z=44$ and $Z=40$ respectively).  
  The expectation is that the two systems will have the {\em same} backgrounds while noticeably {\em different} CME signals   due to the difference in their  nuclear charge and thus magnetic field strength. This experiment offers the unique opportunity to detect CME in such collisions and the data analysis is actively underway.  A precise characterization of the  signals and backgrounds is critically needed.

 The present work focuses on making theoretical predictions for the signatures of CME in the isobar experiment. For that purpose, we develop a  state-of-the-art tool and the first of its kind, the EBE-AVFD (event-by-event anomalous-viscous fluid dynamics), that can characterize  CME signals from dynamical anomalous transport as well as account for background correlations in a realistic heavy ion collision environment. With this powerful tool, we compute  CME observables and report key results that shall provide unique  insights into forthcoming experimental measurements.

  
{\em Methodology.---} In heavy ion collisions, the Chiral Magnetic Effect induces an electric current along the magnetic field, approximately perpendicular to the reaction plane (RP)~\cite{Bloczynski:2012en}. Therefore a CME-induced charge separation across the reaction plane is expected  and can  be measured by the charge asymmetry in azimuthal correlations of same-sign (SS) and opposite-sign (OS) charged hadron pairs. There are however non-CME background  correlations  contributing substantially to relevant observables, with resonance decays and local charge conservation  being dominant sources. To unambiguously extract the CME signal  has proven extremely challenging.  To resolve such pressing issue and pave the way for potential discovery of CME would require: (1) a sophisticated and realistic simulation framework that can quantitatively characterize backgrounds and predict the CME signatures; (2) a model-independent analysis approach to subtract out backgrounds. The methodology we adopt in this study aims precisely to address these, by developing the EBE-AVFD framework and by a suitable isobar comparison strategy.


To quantitatively describe CME-induced signatures in the collisions, one needs to compute the anomalous charge transport current in dynamically evolving bulk fluid modeled by relativistic viscous fluid dynamics~\cite{Shen:2014vra,Gale:2013da}.  (We note that there are efforts  in simulating CME based on non-hydrodynamic models~\cite{Deng:2016knn,Sun:2018idn,Zhao:2019crj}.)   
Based on theoretical foundation from \cite{Son:2009tf}, we present here a full-fledged fluid dynamical realization  of CME transport in modeling heavy ion collisions. 
This new framework is built upon our earlier Anomalous-Viscous Fluid Dynamics (AVFD)~\cite{Shi:2017cpu,Jiang:2016wve}, which describes the  dynamical evolution of fermion currents (i.e. the quark currents of various flavors and chirality)  perturbatively  on top of the neutral bulk fluid. 
A few crucial elements however were missing, including background correlation implementations, event-by-event fluctuations and a hadron cascade stage after hydro-stage.

In the present work, we've successfully addressed these  challenges by developing the Event-By-Event Anomalous-Viscous Fluid Dynamics (EBE-AVFD). In the EBE-AVFD, the initial state fluctuations are fully accounted for by event-wise sampling for bulk entropy density and fermion axial charge density.   Following the end of hydrodynamic stage (using VISHNU hydro simulations) for each bulk event evolution, hadrons are sampled by maintaining charge conservations and then further evolved through hadron cascade stage via URQMD simulations. This framework for the first time allows a quantitative and consistent evaluation of both CME signals and background correlations within the same realistic bulk evolution. The EBE-AVFD represents state-of-the-art tool for  reliable predictions of  CME signatures  in heavy ion collisions. 

The key to the success of the isobaric contrast idea, is to make sure that {\em one has two  collections of RuRu and ZrZr collision events  that must be identical in their bulk properties (multiplicity and elliptic flow $v_2$)}. In conventional analysis  one would select events based on centrality (i.e. multiplicity) and then compare RuRu with ZrZr systems at the same centrality. Recent simulations of initial geometry in these collisions however suggest difference at a few percent level in their elliptic eccentricity (for the same centrality) due to uncertainty in the nucleon distributions of the isobar nuclei~\cite{Xu:2017zcn,Shi:2018sah,Hammelmann:2019vwd}.  This presents enough of concern which may complicate the supposedly ``clean'' isobar comparison, given the small CME signal.  
To ensure a successful isobar contrast, we propose a new strategy for comparing the isobaric systems, namely to  use a joint (multiplicity + elliptic flow) event selection method~\cite{Shi:2018sah}.   
We have extensively verified the effectiveness of this strategy based on joint (multiplicity + elliptic flow) event selection for a variety of nucleon distributions,  
both with event-by-event simulations of initial conditions and   with the final state events from EBE-AVFD simulations. More specifically, the elliptic flow coefficient $v_2$ in our simulations is computed by first identifying event plane $\Psi_{EP}$ with all charged particles within the rapidity range $y\in (1.5,4.0)$ and then evaluating $v_2 =\langle cos(2\phi-2\Psi_{EP}) \rangle $ for charged particles at mid-rapidity $y \in (-1,1)$.
 We find that between the isobar pairs, the relative difference in geometry 
 is reduced to the level of $\sim 0.1\%$ while the relative difference in magnetic field 
remains at the $10\sim25\%$ level. 
Therefore, such an isobar comparison and subtraction strategy allows an unambiguous contrast between RuRu and ZrZr to reveal potential magnetic-field-driven CME signal despite uncertainty about the initial nucleon distributions.  


{\em Predictions.---} Here we present predictions for isobaric collisions from our EBE-AVFD simulations. 
 Focusing on events corresponding to  $40\sim50\%$ centrality range,  we have generated ten millions  of collision events for each of the RuRu and ZrZr systems. For each system, EBE-AVFD simulations are done for $10^5$ different hydrodynamic initial profiles, with $100$ hadron cascade events following each hydrodynamic profile. We treat each hadron-cascade outcome independently  and use all these $10^7$ events together for our analysis of final state observables such as multiplicity, elliptic flow and charge-dependent correlations.
We apply the identical joint cut for multiplicity $N^{ch}$ and elliptic flow $v_2$:  $65\le N^{ch}_{|y|<1} \le 96$ and $0.05 < v_2 < 0.25$ for both systems.  
We have done a consistency check by comparing the post-selection events from both RuRu and ZrZr which are 
found to be   {\em identical}, with   $\langle N_{ch} \rangle =80.4$ (standard deviation $8.5$) and  $\langle v_2 \rangle =0.1132$ (standard deviation $0.046$). This procedure thus guarantees the same bulk medium and   background correlations for the isobar pairs.

With the selected events we perform analysis of CME-motivated observables, focusing on the absolute difference between RuRu and ZrZr systems which would subtract out the background portion. What potentially remains from such subtraction would be the pure CME signal.  
This allows revealing characteristic features  expected for a pure CME signal (obtained only after removal of backgrounds via isobar subtraction). The strength of such a pure signal   shall be quadratically dependent on the  initial axial charge. The correlations from such a signal shall be along the magnetic field direction which would have different degrees of correlation with the event-plane (EP) and the reaction-plane (RP)~\cite{Bloczynski:2012en,McLerran:2013hla,Gursoy:2014aka,Tuchin:2015oka,Inghirami:2016iru,Gursoy:2018yai,Roy:2017yvg,Pu:2016ayh,Muller:2018ibh,Guo:2019joy,Guo:2019mgh}.  
These features would provide key validation of the EBE-AVFD predictions for isobaric collisions.


The CME current leads to a {\em charge separation} along the magnetic field direction that can  be measured with  azimuthal correlations such as the $\gamma$-correlator~\cite{Voloshin:2004vk}: 
\begin{eqnarray} \label{eq_gamma}
\gamma^{\alpha\beta}=\langle \cos \left( \phi^\alpha + \phi^\beta - 2\Psi_{2} \right) \rangle 
\end{eqnarray}
where $\Psi_2$ should ideally be the reaction plane (RP) but  is practically identified via the event plane (EP). 
The correlation of pairs with same electric charge, $\gamma^{SS}$, has $\{ \alpha \beta \} \to \{ ++\}$ or $\{ -- \}$ while that for opposite charged pairs,  $\gamma^{OS}$, has $\{ \alpha \beta \} \to \{ +- \}$ or $\{ -+ \}$. 
To maximize the signal and reduce the backgrounds, one can further examine the difference between the correlation of same and opposite charged pairs, $ \gamma^{OS-SS}=\gamma^{OS}-\gamma^{SS} $.   Another closely related observable is the $\delta$-correlator~\cite{Bzdak:2012ia,Bzdak:2009fc}
\begin{eqnarray}\label{eq:jl:delta}
\delta^{\alpha \beta} = {\big \langle} \cos\left(\phi^\alpha - \phi^\beta  \right) {\big \rangle} \,\, .
\end{eqnarray}  
with $ \delta^{OS-SS}=\delta^{OS}-\delta^{SS} $.

The CME  would contribute to the above correlators as : $\gamma^{OS-SS}_{CME} \to 2 \langle a_1^2 \cos\left(2\Psi_B - 2\Psi_2 \right) \rangle $ and 
$\delta^{OS-SS}_{CME} \to -2 \langle a_1^2   \rangle $ where $a_1$ is the event-wise charge separation dipole   along the magnetic field direction $\Psi_B$ with $\cos\left(2\Psi_B - 2\Psi_2 \right)$ capturing the azimuthal de-correlation between $\Psi_B$ and $\Psi_2$. However, CME contributions could not be easily extracted from current measurements of $\gamma$ and $\delta$ correlators due to dominant non-CME backgrounds~\cite{Bzdak:2009fc,Schlichting:2010qia,Bloczynski:2013mca,Wen:2016zic,Xu:2017qfs,Zhao:2017nfq,Huang:2019vfy,Choudhury:2019ctw}.  This is where the isobar contrast would be uniquely valuable. We will focus on their difference: $\gamma^{OS-SS}_{Ru-Zr} = \gamma^{OS-SS}_{RuRu} -  \gamma^{OS-SS}_{ZrZr}$ and $\delta ^{OS-SS}_{Ru-Zr} = \delta^{OS-SS}_{RuRu} -  \delta^{OS-SS}_{ZrZr}$.

\begin{figure}[htb!]
\includegraphics[width=6.6cm]{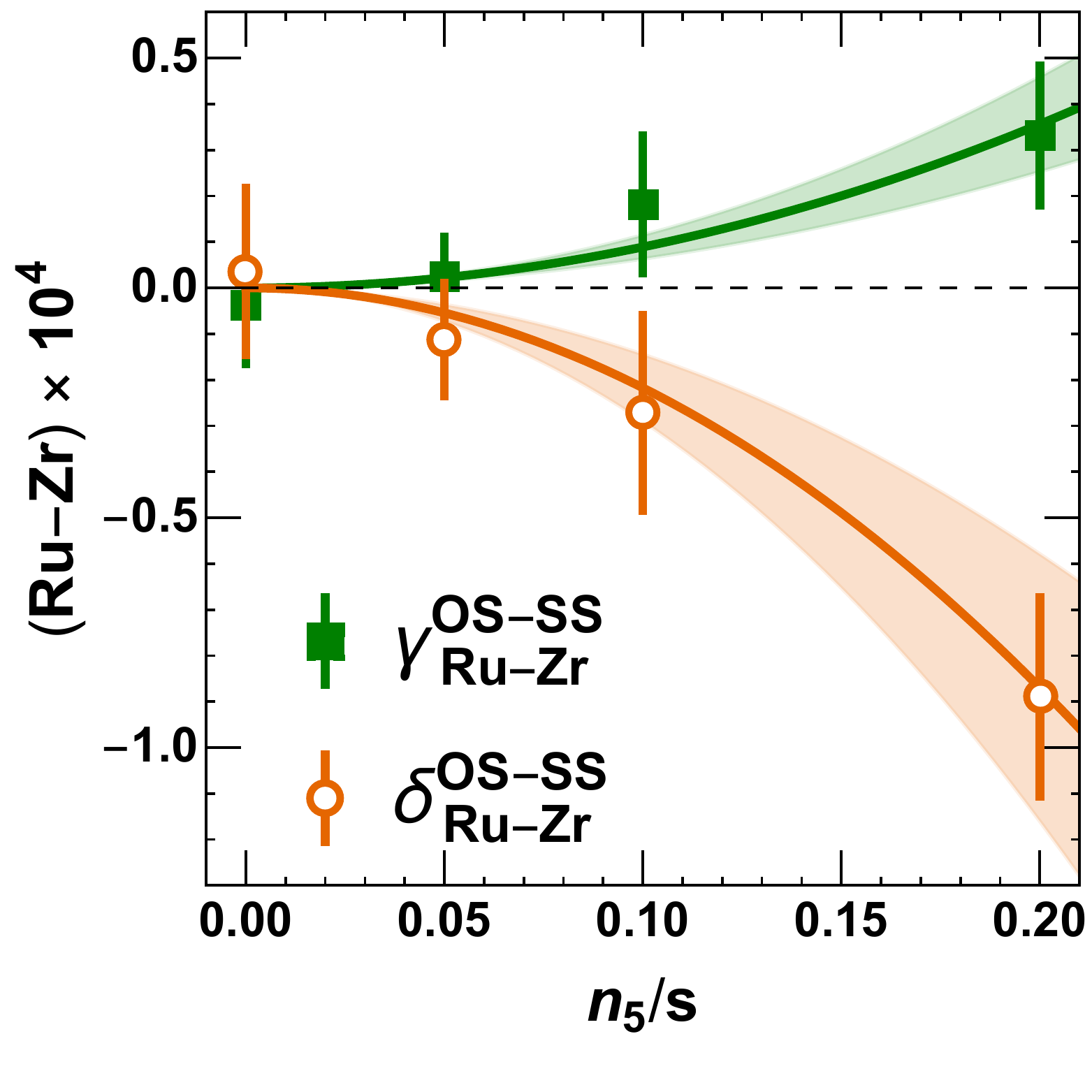}\vspace{-0.2in}
\caption{\label{fig_EP} (color online) 
EBE-AVFD predictions for $\gamma_{Ru-Zr}^{OS-SS}$ (solid square) and $\delta_{Ru-Zr}^{OS-SS}$ (open circle) with respect to event-plane (EP) (i.e. $\Psi_2 \to \Psi_{EP}$) for $n_5/s = 0\%, 5\%, 10\%$ and $20\%$ respectively.  Vertical error bars are statistical uncertainty from simulations. The curves are quadratic fitting results with shaded uncertainty bands.} \vspace{-0.08in}
\end{figure}

In Fig.~\ref{fig_EP}  we show  EBE-AVFD predictions   for   $\gamma_{Ru-Zr}^{OS-SS}$ and $\delta_{Ru-Zr}^{OS-SS}$ with respect to event-plane (EP) geometry (i.e. $\Psi_2 \to \Psi_{EP}$), for four different levels of initial axial charge density  (normalized by initial bulk entropy density $s$) at $n_5/s = 0\%, 5\%, 10\%$ and $20\%$ respectively.   The quantitative consistency between $\gamma$ and $\delta$ correlators would provide sufficient validation of CME-signal and help constrain the uncertainty in initial axial charge.

The CME-induced correlations should depend quadratically on the $n_5$. Simulation results indeed show such a trend, with quadratic fitting curves presented in Fig.~\ref{fig_EP} and summarized below: 
\begin{eqnarray} 
\gamma^{OS-SS}_{Ru-Zr} {\bigg |_{EP}}  \simeq \    \left (   0.89 \pm 0.51\right ) \times 10^{-3} \times  \left ( \frac{n_5}{s} \right)^2 \\ 
\delta ^{OS-SS}_{Ru-Zr}  {\bigg |_{EP}}  \simeq  \  -\left  ( 2.17 \pm 0.72  \right ) \times 10^{-3} \times  \left ( \frac{n_5}{s} \right)^2
\end{eqnarray}

We further propose a new observable $\zeta_{isobar}$ built from the ratio between the two correlators: 
\begin{eqnarray} \label{eq_zeta_EP}
\zeta_{isobar}^{EP} \equiv  \frac{ \gamma^{OS-SS}_{Ru-Zr} {\bigg |_{EP}}  }{\delta ^{OS-SS}_{Ru-Zr}  {\bigg |_{EP}}} 
\simeq - ( 0.41 \pm  0.27 ) 
\end{eqnarray}
 This new ratio  is {\em independent} of the (uncertain) initial axial charge and therefore provides a robust test of CME. The ratio essentially reflects  the azimuthal de-correlation $\cos\left(2\Psi_B - 2\Psi_{EP} \right)$, which is independently computed to be  about $0.46$ and quantitatively consistent with the above ratio. These features are specific to pure CME signal and are manifested only after isobar subtraction.  


Comparison of measurements with respect to  reaction plane (RP) and to event plane (EP) could help decipher CME signal~\cite{Xu:2017qfs,Voloshin:2018qsm}, as  the magnetic field has different degrees of azimuthal de-correlations with RP and with EP. Experimentally one may use  the spectator plane (from e.g. ZDC) as a proxy for RP. It would be interesting  to examine {\em correlators with respect to RP}. 

\begin{figure}[htb!]
\includegraphics[width=6.6cm]{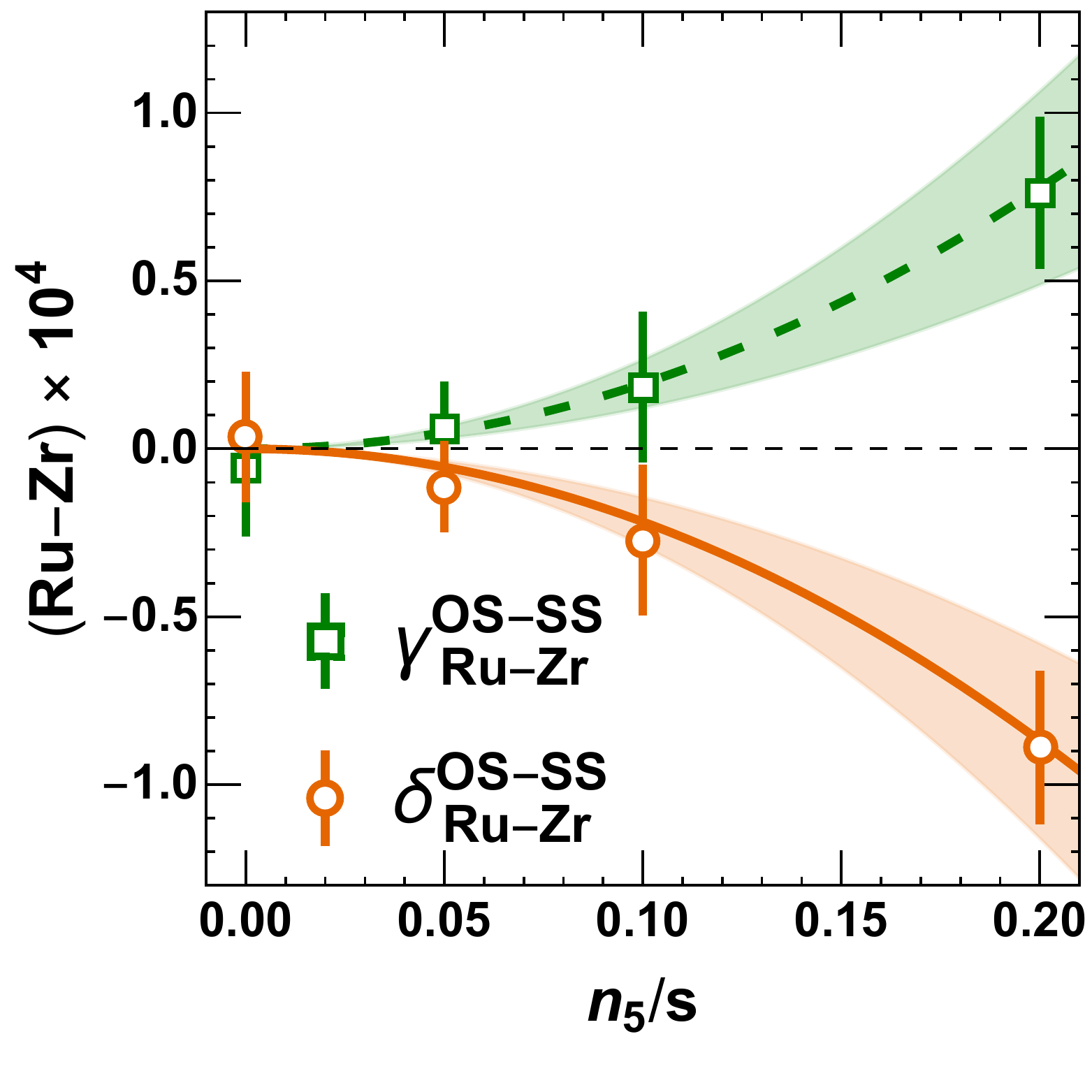}\vspace{-0.2in}
\caption{\label{fig_RP} (color online) 
Same as Fig.~\ref{fig_EP} but for correlations measured with respect to the reaction-plane (RP)  (i.e. $\Psi_2 \to \Psi_{RP}$).}\vspace{-0.2in}
\end{figure}

In Fig.~\ref{fig_RP}  we show EBE-AVFD predictions   for   $\gamma_{Ru-Zr}^{OS-SS}$ and $\delta_{Ru-Zr}^{OS-SS}$ with respect to reaction-plane  (i.e. $\Psi_2 \to \Psi_{RP}$).  Compared with EP results in Fig.~\ref{fig_EP}, $\gamma$ correlator becomes larger due to a stronger correlation between magnetic field and the RP, as uniquely expected for pure CME signal.   Fig.~\ref{fig_RP} also presents quadratic fitting curves: 
\begin{eqnarray} 
\gamma^{OS-SS}_{Ru-Zr} {\bigg |_{RP}}  \simeq \    \left (   1.94 \pm 0.72 \right ) \times 10^{-3} \times  \left ( \frac{n_5}{s} \right)^2 \\ 
\delta ^{OS-SS}_{Ru-Zr}  {\bigg |_{RP}}  \simeq  \  -\left  ( 2.17 \pm 0.72  \right ) \times 10^{-3} \times  \left ( \frac{n_5}{s} \right)^2
\end{eqnarray}  
The RP result for the ratio  observable $\zeta_{isobar}$ is:  
\begin{eqnarray}
\zeta_{isobar}^{RP} \equiv  \frac{ \gamma^{OS-SS}_{Ru-Zr} {\bigg |_{RP}}  }{\delta ^{OS-SS}_{Ru-Zr}  {\bigg |_{RP}}} 
\simeq - (  0.90 \pm 0.45 ) 
\end{eqnarray}
This RP  ratio  is about twice that from EP, in quantitative consistency with the expected   de-correlation factor $\cos\left(2\Psi_B - 2\Psi_{RP} \right)$ of about $0.95$. The EP and RP measurements of these correlators and the proposed ratios would together provide a stringent test  for validating the CME signatures.

\begin{figure}[htb!]
\includegraphics[width=6.cm]{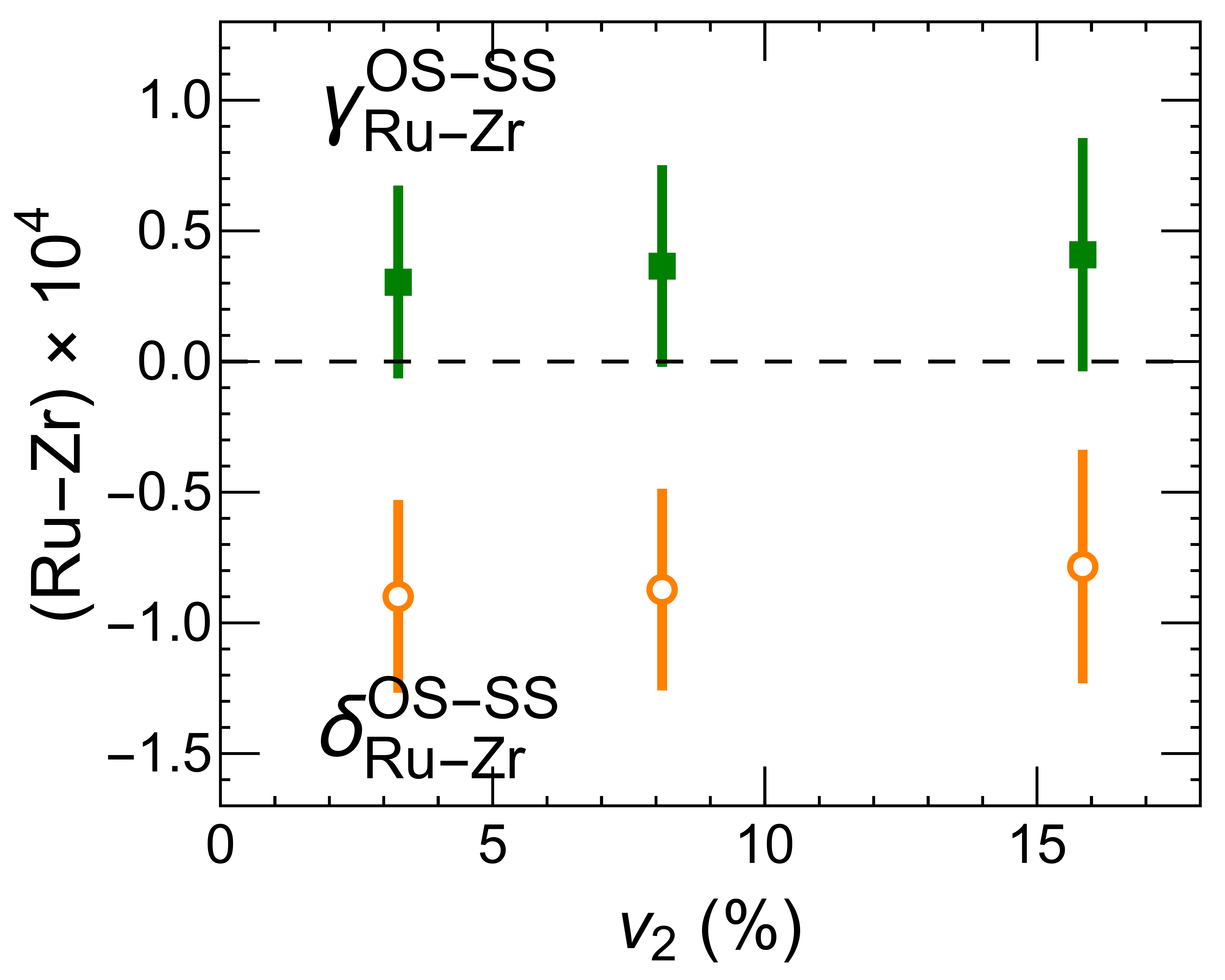}\vspace{-0.1in}
\caption{\label{fig_ESE} (color online) 
EBE-AVFD predictions for observables $\gamma_{Ru-Zr}^{OS-SS}$ and $\delta_{Ru-Zr}^{OS-SS}$ as a function of bin-wise elliptic flow $v_2$  from event-shape analysis with  three identical bins for RuRu and ZrZr systems. The simulation results are obtained with   $n_5/s =  20\%$. }\vspace{-0.2in}
\end{figure}

Event-shape analysis provides a way of revealing the backgrounds, showing the dependence of $\gamma$-correlator on bulk $v_2$~\cite{Acharya:2017fau,Sirunyan:2017quh}.  A pure CME signal, on the other hand, should be (nearly) independent of event shape. This provides an important consistency check for CME signal from isobar subtraction. In Fig.~\ref{fig_ESE} we show EBE-AVFD results for   $\gamma_{Ru-Zr}^{OS-SS}$ and $\delta_{Ru-Zr}^{OS-SS}$ versus event shape in  three bins: $v_2\in (0.01, 0.055)$,  $v_2\in (0.055, 0.11)$ and   $v_2\in (0.11, 0.30)$. 
 We indeed observe that isobar-subtracted $\gamma$ and $\delta$ are independent of event shape.

\begin{figure}[htb!]
\includegraphics[width=7.cm]{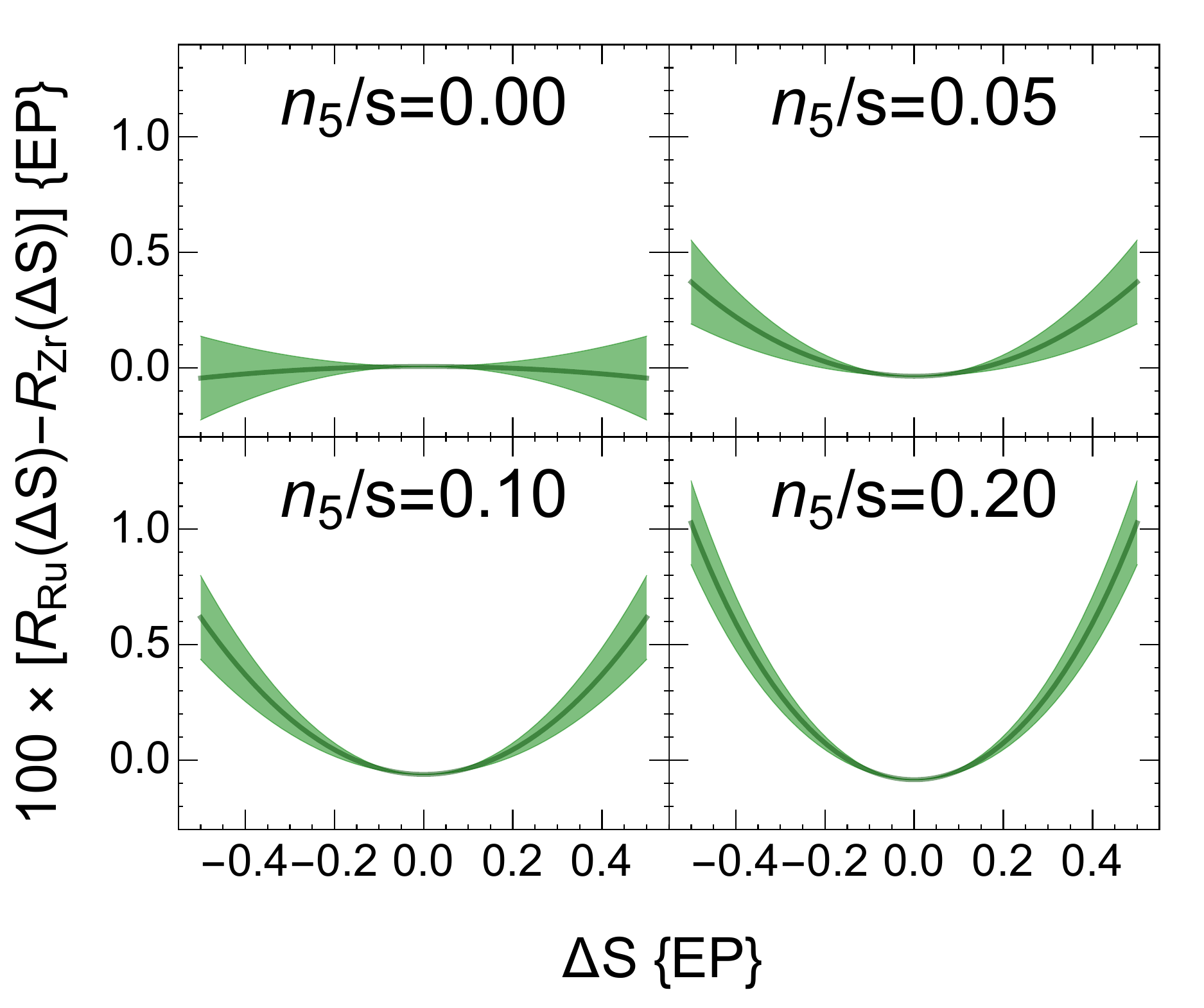}\vspace{-0.1in}
\caption{\label{fig_RC} (color online) 
EBE-AVFD predictions   for the R-correlator distributions for $n_5/s = 0\%, 5\%, 10\%$ and $20\%$ respectively. }\vspace{-0.1in}
\end{figure}


A number of other CME-sensitive correlators have also been proposed~\cite{Magdy:2017yje,Magdy:2018lwk,Tang:2019pbl}. As an example,    the so-called  {\em R-correlator} (---see  \cite{Magdy:2017yje,Magdy:2018lwk} for detailed definition and discussions)  has demonstrated a certain sensitivity to the presence of CME.  In Fig.~\ref{fig_RC} we show the EBE-AVFD results for the  the isobar-subtracted R-correlator, $[R_{Ru} (\Delta S) - R_{Zr} (\Delta S)]$  for $n_5/s = 0\%, 5\%, 10\%$ and $20\%$ respectively. It appears  flat for the none-CME case ($n_5/s=0\% $) while becomes more and more upward concave with increasing  CME signal. Measurement of R-correlator with enough statistics should provide further   validation of the CME signal.


{\em Conclusion.---} In this work we've presented a comprehensive set of quantitative predictions for the signatures of Chiral Magnetic Effect (CME) in the isobaric collisions at RHIC. Reliable predictions for key observables of such a decisive high-profile experiment is critically needed and becomes possible only with the development of the EBE-AVFD framework reported in the present study. This novel tool has allowed us to characterize a number of unique features of CME signals in these collisions and to propose the best strategy for an unambiguous validation of CME with a multitude of measurements. These results represent the state-of-the-art understanding of the problem, offers valuable insights for the ongoing data analysis and interpretation, and  would help significantly advance the current search of CME in collider experiments. 

Given the predicted signal strength and the currently projected measurement precision, we conclude optimistically about the likelihood for a successful extraction of CME signatures in the collisions of isobars. If confirmed indeed, such a detection  would not only be the first observation of CME in a subatomic material with intrinsically relativistic chiral fermions, but also provide the tantalizing evidence of QCD chiral symmetry restoration in the quark-gluon plasma as well as the unique manifestation of the elusive QCD gluon topological fluctuations. Establishing the CME phenomenon in the subatomic chiral matter would also have far-reaching implications for the study of CME in other areas of research such as condensed matter physics.

\section*{Acknowledgments}
%
This work is supported in part by the NSFC Grants No. 11735007 and No. 11875178, by the NSF Grant No. PHY-1913729 and by the U.S. Department of Energy, Office of Science, Office of Nuclear Physics, within the framework of the Beam Energy Scan Theory (BEST) Topical Collaboration.  SS is grateful to the Natural Sciences and Engineering Research Council of Canada for support.  The computation of this research was performed on IU's Big Red II cluster, 
that is supported by Lilly Endowment, Inc., through its support for the Indiana University Pervasive Technology Institute, and   by the Indiana METACyt Initiative which  was also partly supported  by Lilly Endowment, Inc.


%
%
%
\bibliography{Isobar.bib} 



%

\end{document}